\documentclass[twocolumn,aps,prb,showpacs]{revtex4-1}
\usepackage[latin1]{inputenc}
\usepackage{amsmath}
\usepackage{amsfonts}
\usepackage{graphicx}
\usepackage{exscale}
\usepackage{amsbsy}
\usepackage{amssymb}
\usepackage{makeidx}

\begin{document}

\title{\bf Correlations and Renormalization of the Electron-Phonon Coupling in the Honeycomb Hubbard Ladder and Superconductivity in Polyacene }

\author{G. Karakonstantakis${}^{1}$}
\author{L. Liu${}^{1}$}
\author{R. Thomale${}^{2,3}$}
\author{S. A. Kivelson${}^{1}$}

\affiliation{${}^{1}$Department of Physics, Stanford University, Stanford, California 94305, USA.}
\affiliation{${}^{2}$Institut de th\'eorie des ph\'enom\`enes physiques, \'Ecole Polytechnique F\'ed\'erale de Lausanne (EPFL), CH-1015 Lausanne}
\affiliation{${}^{3}$Institute for Theoretical Physics and Astrophysics, University of W\"urzburg, D 97074 W\"urzburg}

\newcommand{\bs}[1]{\boldsymbol{#1}}
\newcommand{\comm}[2]{\left[#1,#2\right]}
\newcommand{\anticomm}[2]{\left\{#1,#2\right\}}
\newcommand{\vac}{\left|\,0\,\right\rangle}

\newcommand\avg[1]{\left\langle#1\right\rangle}
\newcommand\bra[1]{\left\langle#1\right|}
\newcommand\ket[1]{\left|#1\right\rangle}
\newcommand\braket[2]{\left\langle #1\right|\left.#2\right\rangle}
\newcommand\combin[2]{\kc{^{#1}_{#2}}}
\newcommand\abs[1]{\left|#1\right|}
\newcommand\kc[1]{\left(#1\right)}
\newcommand\kd[1]{\left[#1\right]}
\newcommand\ke[1]{\left\{#1\right\}} 
\renewcommand\Re{\rm Re}
\renewcommand\Im{\rm Im}
\newcommand\sgn{\rm sgn}
\newcommand\mode{\text{ }{\rm mod}\text{ }}
\newcommand\trace[1]{{\rm Tr}\left[#1\right]}
\newcommand\Det[1]{{\rm Det}\left(#1\right)}
\newcommand\be{\begin{equation}}       
\newcommand\ee{\end{equation}}
\newcommand\bea{\begin{eqnarray}}      
\newcommand\eea{\end{eqnarray}}
\newcommand\ba{\begin{array} }
\newcommand\ea{\end{array} }
\newcommand\bnum{\begin{enumerate} }
\newcommand\enum{\end{enumerate}}
\newcommand\lf{\left}
\newcommand\rt{\right}
\newcommand\nn{\nonumber}
\newcommand\pa{\partial}
\newcommand\Ra{\Rightarrow}
\newcommand\ra{\rightarrow}
\newcommand\A{\uparrow}
\newcommand\V{\downarrow}
\newcommand\PRB{Phys. Rev. B}
\newcommand\PRL{Phys. Rev. Lett.}
\newcommand\vk{{\bf k}}
\newcommand\Tau{Fathcal{T}}
\newcommand\Ep{Fathcal{E}}
\newcommand\Kp{Fathcal{K}}
\newcommand\intz{Fathbb{Z}}
\newcommand\vect[1]{\kc{\ba{c}#1\ea}}
\newcommand\elist[1]{\left\{\ba{cc} #1\ea\right.}
\newcommand\Cl{{\rm Cliff}}
\newcommand\Pf[1]{{\rm Pf}\kc{#1}}
\newcommand\m{\textrm{matter}}
\newcommand\g{\textrm{gauge}}
\newcommand\Eq[1]{Eq.~(\ref{#1})}
\newcommand\Fig[1]{Fig. ~\ref{#1}}
\newcommand\te{\mathrm{e}}
\newcommand\eff{\mathrm{eff}}
\newcommand\tr{\mathrm{tr}}
\newcommand\Tr{\mathrm{Tr}}
\newcommand\bS{{\bf S}}
\newcommand\tS{{\tilde S}}
\newcommand\tP{{\tilde P}}

\begin{abstract}

We have performed extensive density matrix renormalization group (DMRG) studies of the Hubbard model on a honeycomb ladder.  The band structure (with Hubbard $U=0$) exhibits an unusual quadratic band touching at half filling, which is associated with a quantum Lifshitz transition from a band insulator to a metal.
For one electron per site, non-zero $U$ drives the system into an insulating state in which there is no pair-binding between added electrons;  this implies that  superconductivity driven directly by the repulsive electron-electron interactions is unlikely in the regime of small doping, $x\ll 1$.  However, the divergent density of states as $x\to 0$, the large values of the phonon frequencies, and an unusual correlation induced enhancement of the electron-phonon coupling imply that lightly doped polyacenes, which approximately realize this structure, are good candidates for high temperature electron-phonon driven superconductivity. 

\end{abstract}
\maketitle
\section{Introduction}
To obtain high temperature superconductivity from an electron-phonon mechanism one would like to find a material with a reasonably strong electron-phonon coupling ($\lambda=N(E_{\text{F}})V^{\text{eff}} \sim 1$ where $N(E_{\text{F}})$ is the electronic density of states per unit cell and $V^{\text{eff}}$ is the strength of the induced attraction), and a large phonon frequency, $\omega_0$, to produce  a large prefactor for any BCS-like expression for $T_c$~\cite{bcs}.  However, it is also important that $\hbar\omega_0/E_{\text{F}} \ll 1$, so that the bare repulsive interactions between electrons are significantly renormalized (decreased) by electronic fluctuations with energies between $\hbar\omega_0$ and $E_{\text{F}}$.  Since $V^{\text{eff}}\propto 1/K$, where $K$ is the phonon stiffness constant, while $\omega_0 \propto\sqrt{K}$, there is a general tendency for large phonon frequencies to be associated with weak pairing interactions. However, as $\omega_0 \propto \sqrt{1/M}$, where $M$ is the nuclear mass, while $V^{\text{eff}}$ is independent of $M$, it is  more likely that a substantial $V^{\text{eff}}$ {\em and} a large $\hbar \omega_0$ will be found in materials with light elements.  A substantial $\lambda$ is also more likely in systems with a large value of $N(E_{\text{F}})$ but this either requires a pronounced maximum in $N(E)$ near $E=E_{\text{F}}$,  or a small value of $E_{\text{F}}$;  however, a reduced value of $E_{\text{F}}/\hbar\omega_0$ can eliminate the retardation which is  essential for this  mechanism of superconductivity. 

\begin{figure}[t]
\includegraphics[width=3.2in]{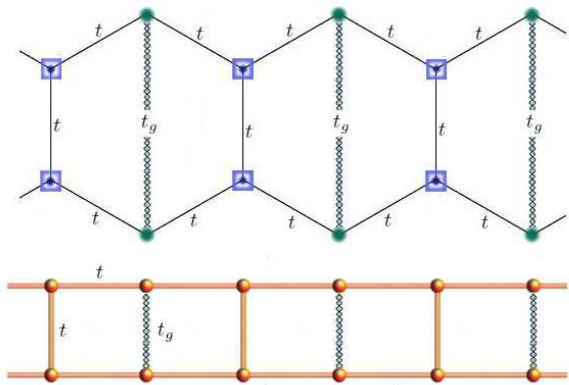}
\caption{(Color online) a) A honeycomb two-leg ladder with the same nearest neighbor hopping amplitude for all
sites with magnitude $t$ and third neighbor hoping amplitude $t_g$ only between the top and the bottom
of each hexagon. This is equivalent to a two-leg ladder 
with alternating rung hopping matrix elements $t$ and $t_g$, as shown in b).
}
\label{fig:ham-ladder}
\end{figure}

As was suggested by one of us\cite{Stevechap}, long polyacene polymers, (which, as shown in 
Fig.~\ref{fig:ham-ladder}a, can be thought of as one-dimensional graphene) would have  many features that are optimal for high temperature superconductivity:  The bandwidths of conducting polymers are large  while the typical phonon frequencies are comparable to those in diamond, but still small compared to $E_{\text{F}}$.  The electron-phonon coupling, however, is only moderate.  For instance, in polyacetylene\cite{ssh} $E_{\text{F}}\approx 5$eV, $\hbar\omega_0\sim 0.1$eV, and $\lambda \approx 0.8$.  However, a peculiarity of the band structure of polyacene leads to a predicted\cite{Stevechap} quadratic band touching at the Fermi energy of undoped polyacene, leading (the system being one-dimensional) to a divergent $N(E_{\text{F}})\sim x^{-1/2}$
where $x$ is the concentration of ``doped electrons,'' or in other words, $1+x$ is the density of electrons per site.

Recently, superconductivity  has been reported in various salts\cite{xue} of relatively short polyacene-like molecules, ranging from 5 to 7 fundamental C hexagons, doped with K.  Although superconductivity in these materials is currently associated with a small minority phase, making the identification of the superconducting species difficult, both the high values of $T_c$ (up to 30K) and the fact that $T_c$ appears to be a strongly increasing function of the number of hexagons in the molecules are encouraging indicators that this is a promising direction for study~\cite{heptacene}.

Here, we report the results of extensive density matrix renormalization group (DMRG)~\cite{white1,white2,scholl} studies of the Hubbard model on the honeycomb ladder in the thermodynamic limit (i.e. extrapolated to  infinite length).  The problem is interesting in its own right, and as a model of the electronic structure of polyacenes.

\begin{figure}[t]
\includegraphics[width=3.2in]{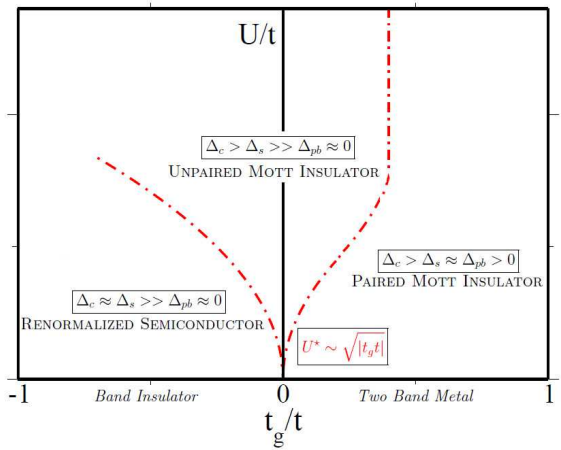}
\caption{(Color online) Schematic phase diagram of the honeycomb 
 ladder, as a function of  $U$ and $t_g$. 
 For $U=0$ there is a two-band metal for $t_g>0$ and a band insulator for $t_g<0$.
For small non-zero $U$ the band insulator evolves into  a renormalized semiconductor
with charge and spin gaps, $\Delta_c \sim \Delta_s$, and the effective interaction between quasiparticles is repulsive (i.e. the pair-binding energy $\Delta_{\text{pb}}\approx 0$). The two-band metal,
evolves into a ``d-Mott insulator,'' in which there is a tendency for pairing upon dopping
({\it i.e.} $\Delta_{\text{pb}}>0$). 
Between these lies a ``fan-shaped'' intermediate regime which collapses to the point $t_g=0$ in the limit $U\to 0$, which is a Mott insulating state (in the sense that $\Delta_c\gg\Delta_s$) with no pairing tendencies ($\Delta_{\text{pb}}\approx 0$).}
\label{fig:phase-diagram}
\end{figure}

Abstractly, the honeycomb ladder  is interesting in that for $U\to0$ it lies precisely at a  Lifshitz quantum critical point.  Fig.~\ref{fig:ham-ladder}a shows a two-leg  Hubbard ladder with alternating hopping matrix elements $t$ and $t_g$ along the rungs;  for $t_g=t$, this is the usual, widely studied two-leg ladder,\cite{2LegHubbard} while for $t_g=0$ it is equivalent to the honeycomb ladder.  A schematic ground state phase diagram of this system extracted from the present study is shown in Fig.~\ref{fig:phase-diagram}.  In the non-interacting limit $U=0$ and  with density of electrons per site $\langle n \rangle=1$, the  system undergoes a  transition from a two-band metal for $t_g > 0$ to a band insulator for $t_g <0$.  For $U>0$, we find that the system is insulating (has a charge gap, $\Delta_c$ 
 - defined in Eq.~\ref{charge}) for all values of $t_g$ and has no phase transitions as a function of $t_g$.  However,  there are   distinct regimes with quite different physical properties separated by crossover lines, as shown in Fig.~\ref{fig:phase-diagram}.    

\noindent{\bf i)}  \underline{ For $0< t_g\lesssim t$} the ground state of the two leg ladder can be thought of as a Mott insulating state of Cooper pairs - a ``paired Mott insulator.''  It is well known\cite{balentsandfisher} that  $U$ is only marginally relevant, so  the charge gap vanishes extremely rapidly,  $\Delta_c \sim t \exp[-1/N(E_{\text{F}}) U]$, as $U/t\to 0$.  Moreover,  the lowest energy state with two added electrons or holes is a charge $2e$ spin-zero bound state with a  pair-binding energy, $\Delta_{\text{pb}}$, comparable to the spin gap, $\Delta_{\text{pb}}\sim\Delta_s$
(defined in Eqs. \ref{PB} and \ref{spingap}, respectively).  

\begin{figure}[t]
\includegraphics[width=2.5in,angle=-90]{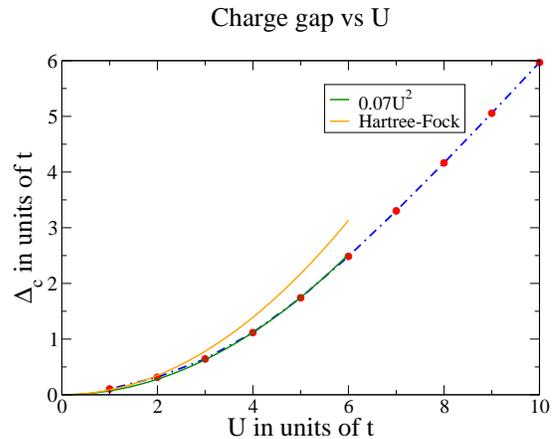}
\caption{(Color online) Charge gap of the undoped honeycomb ladder calculated with DMRG and extrapolated to the thermodynamic limit 
for different $U$. The red dots are the data points. The green line is a quadratic fit
for the first 6 data points, 
a quadratic 
form $\Delta_c\approx 0.07\ U^2/t$, while the yellow line is the leading term of the Hartree-Fock
result, 
$\Delta_{\text{HF}} =0.87\ U^2/t$.}
\label{fig:Uchargegap}
\end{figure}

\begin{figure}[t]
\includegraphics[width=2.5in,angle=-90]{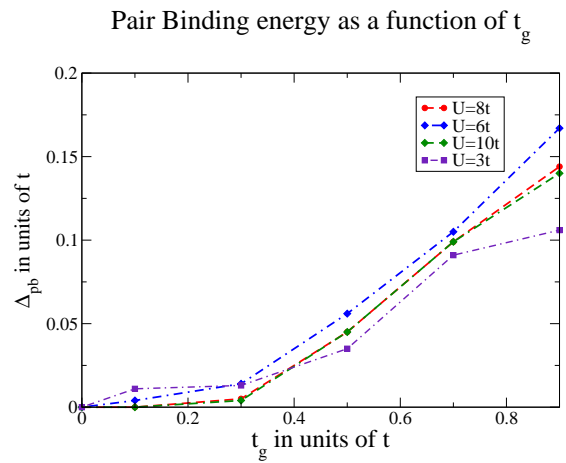}
\caption{(Color online) Pair binding enegry $\Delta_{\text{pb}}$  calculated with DMRG and extrapolated to the thermodynamic limit 
for different $U$. There is  finite $\Delta_{\text{pb}}$ for $t_g >0.3t$, while for $t_g <0.3t$ 
$\Delta_{\text{pb}}$ is vanishingly zero within our numerical uncertaintly. In this specific
calculation, our numerical accuracy is $\pm 0.01t$
}
\label{fig:pairbinding}
\end{figure}

\noindent{\bf ii)}  \underline { For $t_g=0$} and in the non-interacting limit, 
the honeycomb ladder 
corresponds to a quantum critical point with dynamical exponent $z=2$. Correspondingly  we find that $\Delta_c \sim U^2/t$ for small $U$, as shown in Fig.~\ref{fig:Uchargegap}.  Moreover, as is apparent from Fig.~\ref{fig:pairbinding}, for any substantial $U/t$, the pair-binding energy vanishes (or becomes extremely small compared to $\Delta_s$) for $t_g \lesssim 0.3t$, so that in this respect, even though the insulating gap is itself a correlation effect, in terms of the character of the elementary excitations, the system behaves like a conventional semiconductor.
 We refer to the regime that fans out from the point $t_g=U=0$ as an ``unpaired Mott insulator.''  

\noindent{\bf iii)} 
\underline{For $t_g <0$}, $\Delta_c\to |t_g|$ as $U\to 0$, the system is a band insulator with $\Delta_c=\Delta_s >0$ and $\Delta_{\text{pb}}=0$.

The paper is organized as follows. In Section~\ref{sec:model}, we introduce the Hubbard model on the honeycomb ladder and discuss its basic properties such as the quadratic band touching for the non-interacting limit. In Section~\ref{sec:half}, we investigate the half-filled honeycomb Hubbard ladder using Hartree Fock mean field and DMRG approaches. The latter methodology is particularly useful in Section~\ref{sec:nonhalf} where the case away from half filling is studied in detail. Section~\ref{sec:phonon} analyzes how the role of phonon modes is affected by Hubbard-induced electronic correlations.
Transferring our insights to the specific physics of polyacenes, we obtain several highly suggestive results that are discussed in
 great length in Section~\ref{sec:discussion} of this paper, which is phrased in such a way that the reader who wishes to particularly read this discussion can jump straight to that point. In a more speculative discussion which we have moved to Appendix~\ref{app:graphene}, we hypothesize two-dimensional graphene and graphite-based structures, in certain limits of parameters, to be described by weakly coupled quasi-1d honeycomb ladders we have discussed in the paper, and as such to be candidates for high temperature phonon-driven superconductors.

\section{The model}
\label{sec:model}
We study the Hubbard model on the two-leg ladder shown in Fig.~\ref{fig:ham-ladder}b:
\be
H=-\sum_\sigma\sum_{<i,j>}t_{ij}\big[c_{i\sigma}^\dagger c_{j\sigma}+ {\rm h. c.}\big] + U\sum_j n_{j\uparrow}n_{j\downarrow},
\ee
where $c_{j\sigma}$ creates an electron with spin polarization $\sigma$ on site $j$, $n_{j\sigma}=c_{j\sigma}^\dagger c_{j\sigma}$, the site index $j=(\alpha,n)$ with $\alpha=\pm 1$ labels the upper and lower legs of the ladder, and $n=1$ to $L$ labels the distance along the ladder.  We study ladders in the range $L=32$ to $64$, and then extrapolate the results to the thermodynamic limit by plotting various physical quantities as a function of $1/L$.  We assume hopping matrix elements only between pairs of nearest-neighbor sites, where $t_{ij}=t$ for all pairs of nearest-neighbor sites other than every second rung hopping, for which $t_{(1,2n),(2,2n)}=t_g$ with $|t_g| \leq t$.  On the honeycomb ladder shown in Fig.~\ref{fig:ham-ladder}a, this choice with $t_g=0$ corresponds to uniform hopping matrix elements $t$ on all nearest-neighbor bonds, and $t_g$ then is a particular third-neighbor hopping.  As has been noted before\cite{Stevechap}, including second neighbor hopping has no qualitative effect on the band structure, so in addition to allowing us to extrapolate between the uniform two-leg ladder and the honeycomb ladder, the inclusion of a small $t_g$ allows us to explore the most important aspects of the band structure of the honeycomb ladder.

The band structure corresponding to the  non-interacting part of the Hamiltonian (shown in Fig.~\ref{fig:ham-ladder}) can readily be derived:
\be
E_{\lambda,\pm}(k) = \frac{\lambda (t+t_g)}2 \pm \sqrt {\left(\frac {t-t_g} 2\right)^2 + 4t^2\cos^2(k)},
\ee
where $-\pi/2<k\leq \pi/2$ is the Bloch wave-vector and $\lambda=\pm 1$ is the parity of the state under interchange of the upper and lower rung.  (We chose units such that the width of an elementary hexagon is 2.) This band structure is particle hole symmetric, so for $\langle n \rangle =1$, all the negative energy states are occupied and all the positive energy states are empty.  

For $|t_g|\ll t$ the states near the Fermi energy occur near the Brillouin zone (BZ) edge, $k=\pi/2$,
\be
E_{\lambda,\lambda}(k)=-\lambda\Big[ t_g - 4tq^2+ {\cal O}(tq^4) +{\cal O}(t_g^2/t)\Big],
\ee
where $q\equiv k-\pi/2$.
Thus, for $t_g>0$, the band structure is that of a two-band metal, with Fermi surfaces at $k_{\text{F}}=\pi/2 \pm \sqrt{t_g/4t}$.  (This situation evolves smoothly to the limit $t_g=t$, where $k_{\text{F}} = \pi/3, 2\pi/3$ - here the BZ is twice as large.)  For $t_g < 0$, by contrast, the system is a direct gap semiconductor with   a gap of size $2|t_g|$.  Right at $t_g=0$, the band structure exhibits a quadratic band touching, i.e. the honeycomb ladder is tuned precisely to the point of a Lifshitz transition, where the quadratic band dispersion relation corresponds to a dynamic exponent, $z=2$.  
	
We will also be interested in the effects of electron-phonon coupling in this model.  Within the so-called frozen phonon approximation, there are two aspects of this problem that can be analyzed separately:  1) A given pattern of ionic distortion produces a change in the parameters in the effective electronic Hamiltonian.  2) The electronic state responds in a generally complicated fashion to these changes.  The former problem can be treated empirically, by reference to other similar problems -- for instance, the hopping matrix element between nearest-neighbor carbons changes  for small changes in the bond length, $\delta \ell$,  as $t_{ij} = t -\alpha (\delta \ell)$, where in polyacetylene\cite{ssh} (in which the bond-lengths are generally similar), $t\approx 2.5$ eV and $\alpha \approx 4$eV/$\AA$.  

We will explore correlation effects on the latter problem.  Specifically, we will consider the changes in the electronic states produced by small perturbations of the form
\be
H_{\text{el-ph}} = -\sum_{<i,j>\sigma}  \delta t_{ij}\Big[ c_{i\sigma}^\dagger c_{j\sigma} + \text{h.c.}\Big] + \sum_{i,\sigma} \delta \epsilon_i n_{i\sigma},
\label{Helph}
\ee
where, in turn, $\delta t_{ij}$ and $\delta\epsilon_j$ are functions of the ionic distortion in various phonon states.

\section{Half filled band}
\label{sec:half}
Fig.~\ref{fig:phase-diagram} shows a schematic ground-state phase diagram of the model in the $t_g-U$ plane with a mean density of one electron per site ($\langle n \rangle =1$), as inferred from the studies reported below.  For $U>0$ there appears to be a single phase, but a sequence of crossovers (shown as dashed lines) mark the boarders of qualitatively different regimes.  At small $U$, the intermediate regime can be associated with the quantum critical fan opening from the non-interacting Lifshitz quantum critical point at $U=t_g=0$.

\begin{figure}[t]
\includegraphics[width=3.2in]{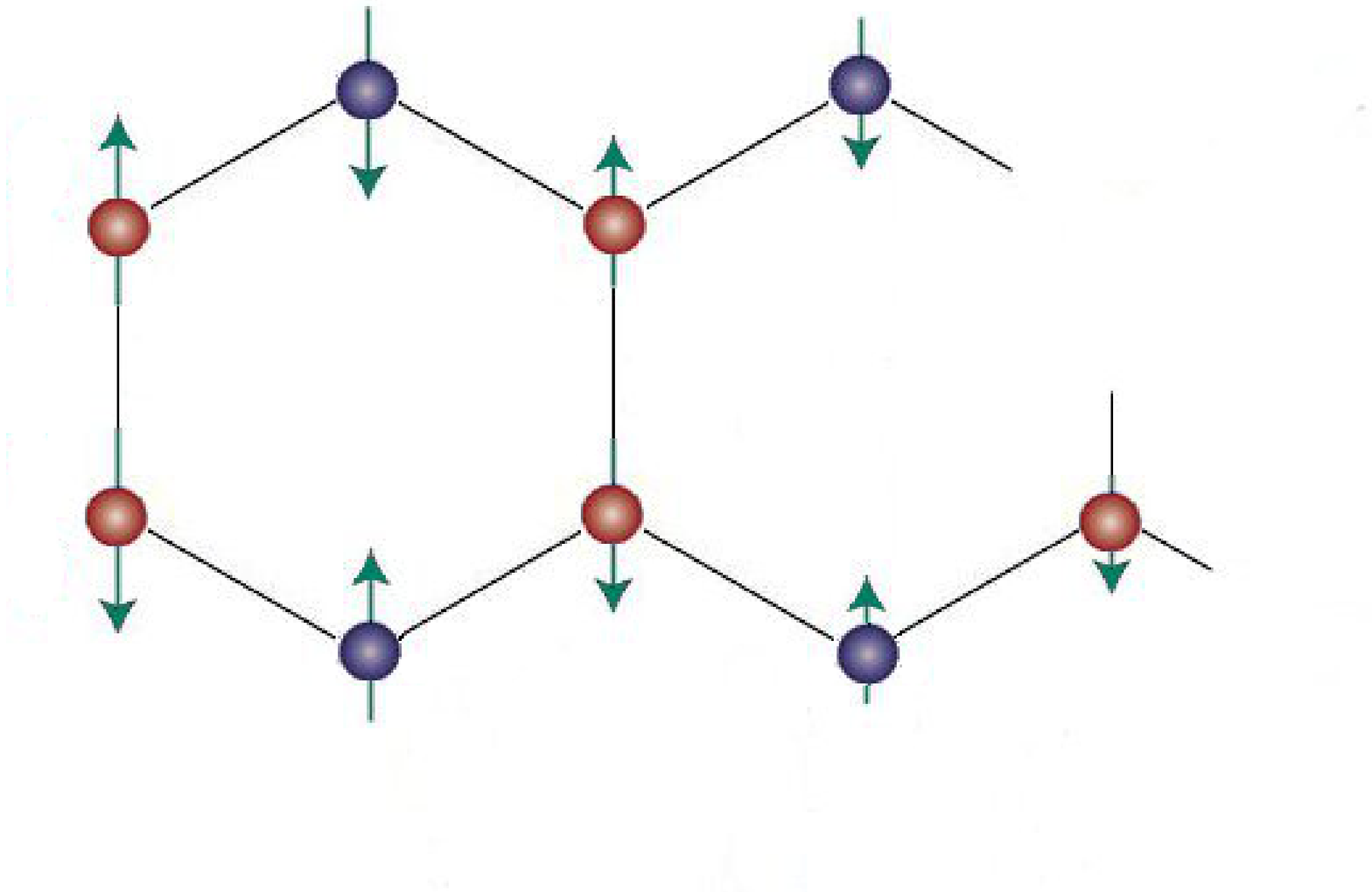}
\caption{(Color online) A schematic illustration of the Hartree-Fock variational ground state. It breaks spin rotation symmetry and the interchangeability of legs.}
\label{fig:Hartree}
\end{figure}

\subsection{Hartree-Fock treatment}
To orient our discussion, we begin by studying the insulating state of the undoped ladder in Hartree-Fock approximation.  For $t_g \geq 0$, where the non-interacting system is metallic, the Hartree-Fock ground-state is insulating and antiferromagnetically ordered, i.e. it spontaneously breaks both spin rotational symmetry and symmetry under interchange of the legs.  The sublattice magnetization is of the form as seen in Fig.~\ref{fig:Hartree}.

\be
\langle \vec S_{(\alpha,i)} \rangle = \bs{\hat{e}} \ m\  (-1)^{i+\alpha}[ 1 + \delta (-1)^i ]
\ee
and the charge density  is of the form
\be
\langle n_{(\alpha,i)} \rangle = 1 + \delta^\prime(-1)^i,
\ee
where the orientation $\bs{\hat e}$ is arbitrary, $m$, $\delta$ and $\delta^\prime$ are  obtained from solving the self-consistency equations, and as such are implicit functions of $U/t$ and $t_g/t$.  $m$ is a monotonically increasing function of $U/t$ with $m=0$ for $U/t=0$ and $m\to 1$ as $U/t\to \infty$.  For any $t_g\neq t$, no symmetry is broken by non-zero values of $\delta$ and $\delta^\prime$, and they do not govern any interesting physics;  they are monotonically decreasing functions of $t_g/t$ and vanish in the limit $t_g/t\to 1$.

For $t_g < 0$, the non-interacting system is gapped, so for small enough $U/t$ no broken symmetry state occurs.  However, for $U> U_c$, the same antiferromagnetic state is recovered.  $U_c$ is a monotonically decreasing function of $t_g/t$, which vanishes as $t_g \to 0$.  

Broken continuous symmetries cannot occur in 1D systems, so the antiferromagnetic order obtained in Hartree-Fock theory is clearly an artifact of the mean-field theory.  Moreover, because there are an integer number of electrons per unit cell, the spin-correlations are expected to decay exponentially with distance at large distances.  

However, Hartree-Fock theory often captures the short-distance physics correctly, and in particular one might expect  the insulating gap to be understandable from these considerations.  For future reference, we define the quasiparticle creation energy to be 
half the energy to create a far separated electron and hole:
\be
\Delta_{\text{qp}}(N) \equiv \frac {{\cal E}_L(2L+1)+{\cal E}_L(2L-1) -2{\cal E}_L(2L)}2,
\label{QP}
\ee
where ${\cal E}_L(N)$ is the  $N$ electron ground-state energy of the $L$ site long ladder (which has $2L$ sites).
 The quasiparticle gap is defined by $\Delta_{\text{qp}}=\lim_{N\to\infty}\Delta_{qp
 }(N)$. In Hartree-Fock theory, $\Delta_{\text{qp}}\sim U m$.  Thus, $\Delta_{\text{qp}} \to U/2$ as $U\to \infty$, independent of $t_g$.  However, for small $U$, different behaviors are seen depending on the value of $t_g$.  For $t_g < 0$, $\Delta_{\text{qp}} = |t_g|$ for small enough $U$.  For $t_g=0$, and small $U$, $m=A^2(U/t)$,
\be
\Delta_{\text{qp}} = A^2U^2/t[1+{\cal O}(U/t)],
\ee
where 
\be
A=\frac{3}{8 \pi}\int dk(\sqrt{1+k^4}-k^2) \approx 0.087.
\ee
For $t_g >0$, the usual considerations of 1D band structures apply, and $m \sim t \exp[ - 1/UN(E_{\text{F}})]$, where, however, $N(E_{\text{F}})$ diverges as $t_g \to 0$ as $N(E_{\text{F}}) \sim 1/\sqrt{tt_g}$.  

There is another feature of the Hartree-Fock solution that one might expect to survive fluctuation effects.  Since in the antiferromagnetic phase there are gapless collective spin-carrying modes, one would expect that even when the long-range order is destroyed by quantum fluctuations, the gap to spin-excitations will be small compared to that to $\Delta_c$.  In particular,  we define the spin gap according to
\be
\Delta_s(N)\equiv \frac{ [{\cal E}_L(2L;S=1) - {\cal E}_L(2L;S=0)]}2,
\label{spingap}
\ee
where ${\cal E}_N(M,S)$ is the ground state energy of the $N$ site long ladder with $M$ electrons and total spin $S$, and again $\Delta_s = \lim_{N\to\infty}\Delta_s(N)$.  Now, although we expect that $\Delta_s>0$ for all $U>0$, we might expect that for $t^\prime \geq 0$ or for $t_g < 0$ but $U \gg U_c$, that $\Delta_{\text{qp}}\gg \Delta_s$, while for $t_g < 0$ and $U<U_c$, $\Delta_{\text{qp}} \approx \Delta_s$.  (Manifestly, for $U=0$, $\Delta_s=\Delta_{\text{qp}}$.)  All these expected features are summarized in the schematic phase diagram in Fig.~\ref{fig:phase-diagram}.

\subsection{DMRG results}

It is possible to compute accurate ground state properties of relatively long two-leg ladders using DMRG.  In most cases, the results can then be straightforwardly and confidently extrapolated to the thermodynamic limit.  
The finite size scaling is carried out using the procedure described in the Appendix of Ref.~\onlinecite{george1}.

To begin with, we compute the various energy gaps.  In addition to the quasiparticle and spin gaps defined in Eqs. \ref{QP} and \ref{spingap} above, we also compute the pair-binding energy,
\be
\Delta_{\text{pb}}(N)\equiv \frac {2{\cal E}_L(2L+1) - {\cal E}_L(2L) - {\cal E}_L(2L+2)}{2},
\label{PB}
\ee
and the charge gap,
\be
\Delta_c(N) \equiv \frac{{\cal E}_L(2L+2,S=0) -{\cal E}_L(2L,S=0)}2,
\label{charge}
\ee
which is the energy per electron charge to add a singlet pair of electrons.  Under a broad range of circumstances, $\Delta_{\text{pb}} = \Delta_{\text{qp}}-\Delta_c$.  

For a band insulator, $\Delta_{\text{pb}}=0$  and $\Delta_s=\Delta_{\text{qp}}=\Delta_c$ while for a singlet superconductor, $\Delta_c=0$ and $\Delta_s=\Delta_{\text{pb}} = \Delta_{min}$, where $\Delta_{min}$ is the minimum of the (in general anisotropic) superconducting gap on the Fermi surface. For the two-leg ladder in the large $U$ limit $\Delta_c \approx \Delta_{\text{qp}} \gg \Delta_s$ and $\Delta_{\text{pb}} \sim \Delta_s$, although the latter can depend on details of the model - for example, adding a  nearest-neighbor repulsion to the model, $U\gg V  \gg t^2/U$, results in $\Delta_{\text{pb}} = 0$.  We will use the occurrence of a non-vanishing (positive) pair binding energy as a crude diagnostic of a tendency to superconductivity.  We will refer to a state with $\Delta_c \gg \Delta_s$ as a Mott insulator, since the insulating character manifestly arises primarily from the strong repulsive interactions between electrons.  We refer to it as a ``paired Mott insulator'' if $\Delta_{\text{pb}} \sim \Delta_s$, and an ``unpaired Mott insulator'' if $\Delta_{\text{pb}}=0$ (or, more pragmatically, if $\Delta_{\text{pb}} \ll \Delta_s$).

\begin{figure}[t]
\includegraphics[width=2.5in,angle=-90]{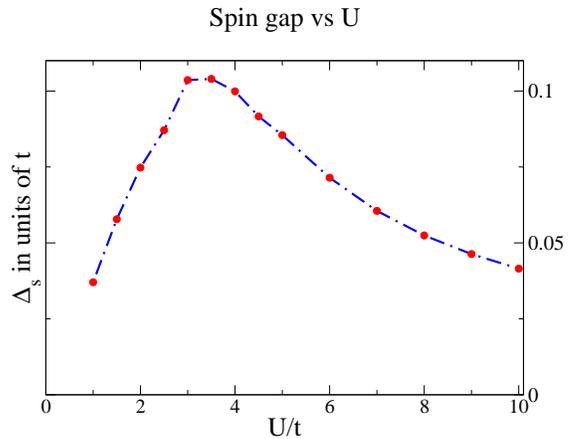}
\caption{(Color online) Spin gap in the thermodynamic limit as a function of $U/t$ for $t_g=0$ at half-filling. There is a 
pronounced maximum for $U=3.5t$ which is roughly the same as the maximum spin gap for the
regular two-leg ladder ($t_g=0$), namely of the size $~0.1t$. }
\label{fig:spingapU}
\end{figure}

\subsubsection{Honeycomb ladder}
Let us first focus on the honeycomb ladder, $t_g =0$.  In Figs~\ref{fig:Uchargegap} and~\ref{fig:spingapU}, we show the charge and spin gaps as a function of $U$.  As expected, $\Delta_c \sim N(E_{\text{F}}) U^2$ for small $U$.  In fact, it is not far from the value predicted by Hartree-Fock theory (dashed line).  The spin-gap is generally much smaller than the charge gap.  We have not been able to extend the DMRG results to values of $U/t<1$, as convergence issues arise.  For the smallest $U$'s we have studied, it appears that $\Delta_s \sim U$, but we suspect that this is simply an intermediate asymptotic of some sort.  The spin-gap has a broad maximum at $U=3t$, where it reaches $\Delta_s = 0.1t$, a value comparable to the  largest spin-gaps found in the two-leg ladder.  However, the pair-binding energy extrapolated to the thermodynamic limit is indistinguishable from zero.  Thus, the honeycomb ladder has an unpaired Mott insulating ground state.

Further  information can be derived from the finite size scaling of various quantities. 
The leading finite size correction to $\Delta_{\text{qp}}$,
\be
\Delta_{\text{qp}}(L) = \Delta_{\text{qp}} + \frac W 2 L^{-2} +{\cal O}(L^{-3}),
\ee
is determined by the quasi-particle effective mass, 
 $W\ \propto\  \hbar^2/m^\star$, 
with a proportionality constant that
depends on the boundary conditions.
In the limit $U\to 0$, for $t=2.5$eV and $a=2.8 \AA$, the value of the effective mass  is 
$m^\star = \hbar^2/(8t)=19.5  \% m_e$ where $m_e$ is the mass of the electron.  
We have found by explicit calculation that $W/t$ does not vary greatly as a function of $U$ 
over the range of $U$ we have explored.
( A value of   $m^\star = 10-20 \% m_e$ has been reported for bilayer graphene\cite{graphene},
comparable to the $U=0$ value we have obtained.)
but the exact relation between $W$ and the effective mass of the electrons $m^\star$
depends on the boundary conditions.

\begin{figure}[t]
\includegraphics[width=2.5in,angle=-90]{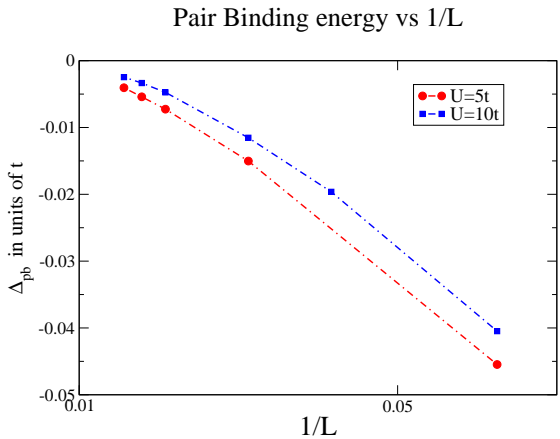}
\caption{(Color online) Length dependence of the pair binding energy for $U=5t$ and $U=10t$ for $L$ up to 64. 
$U^{\text{eff}}$ is obtained from the quadratic fit to the results for $L > 30$.}
\label{fig:scaling}
\end{figure}

The interaction between two quasiparticles can be estimated as well.  If the interaction is relatively weakly repulsive, then there is an intermediate asymptotic satisfied in the perturbative regime $1 \ll N< N^\star$, in which
\be
\Delta_{\text{pb}}(L) = - {3U^{\text{eff}} } L^{-1} + {\cal O}\left( {L^{-2}}\right),
\label{eq:int}
\ee
followed by a long-distance regime, $L > L^\star\sim W/U^{\text{eff}} $ in which
\be
\Delta_{\text{pb}}(L) = -  (3W/2)   L^{-2} +{\cal O}(L^{-3}).
\label{eq:long}
\ee
Independent estimates of $U^{\text{eff}}$ can, in principle, be obtained from studies of systems shorter than $L^\star$ and from the value of $L^\star$ at which a crossover in the scaling behavior occurs.  As shown in Fig. \ref{fig:scaling} for $U=5t$ and $10t$, the length dependence of $\Delta_{\text{pb}}$  can be accurately reproduced by a fit of the form given in Eq.~\ref{eq:int} over the range of $L$ we have explored.  The values of $U^{\text{eff}}$ inferred 
are $U^{\text{eff}} \approx 0.1$ for $U=5t$ and $U^{\text{eff}}\approx 0.2$ for $U= 10t$.  Since these estimates imply $L^\star \sim 100$, which is comparable to or longer than the longest systems we have studied, our analysis is at least self-consistent.

We thus infer an effective model for dilute quasiparticles in the honeycomb ladder:
\be
H^{\text{eff}} = \sum_{k\sigma} E^{\text{eff}}(k) \psi_{k\sigma}^\dagger\psi_{k\sigma}^{\phantom{\dagger}} + U^{\text{eff}}\sum_R\Psi_{R\uparrow}^\dagger\Psi_{R\downarrow}^\dagger\Psi_{R\downarrow}^{\phantom{\dagger}}\Psi_{R\uparrow}^{\phantom{\dagger}},
\label{Heff}
\ee
where $\Psi$ is the Fourier transform of $\psi$, $E^{\text{eff}}(k) = \Delta_{\text{qp}} + \hbar^2k^2/2m^\star + \ldots$ with $m^\star \sim m_e$, $U^{\text{eff}}\sim 0.1 t$ is negligibly small, and $R$ labels the unit cells of the honeycomb ladder.

\begin{figure}[t]
\includegraphics[width=3.2in]{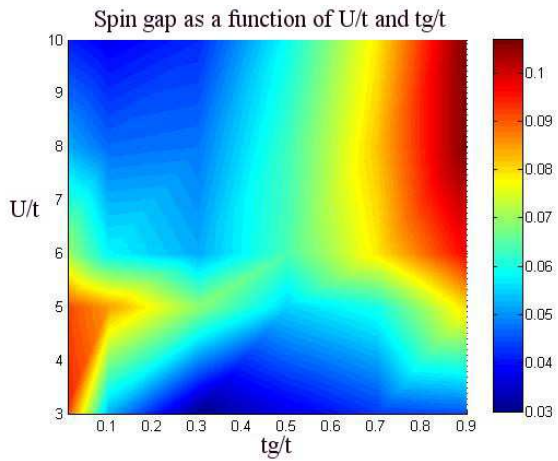}
\caption{
(Color online) Spin gap as a function of  $t_g/t$ 
and
$U/t$. 
The two regions where the spin gap has a local maximum 
are physically distinct:
the upper right side (including the regular two-leg ladder with $t_g=t$) lies in the ``paired-Mott insulating regime'' in which $\Delta_{\text{pb}} \sim \Delta_s$, while  the region on the lower left side (including the pure honeycomb ladder with $t_g=0$) lies in the  "unpaired Mott insulating regime'' where $\Delta_{\text{pb}}\approx 0$.
}
\label{fig:spingap}
\end{figure}

\subsubsection{Non-zero $t_g$}
To see how these results fit in the generalized phase diagram of Fig.  2, we have carried out similar, although less extensive studies of the model as a function of $U$ and $t_g$.  In Fig. \ref{fig:spingap} we show $\Delta_s$ as a contour map.  Note that there are two distinct regimes in which the spin-gap is large --  one for $t_g \approx t$ and $U \sim 8t$, which has been identified in previous studies\cite{george1} as the optimal regime for unconventional pairing in the two-leg ladder, and the other for $t_g \ll t$ and $U \sim 4t$, which includes the honeycomb ladder.  It is clear from this figure that the two regimes are distinct, and that the physics behind the spin-gap formation is likely not the same in the two regimes.  In Fig. \ref{fig:pairbinding} we show $\Delta_{\text{pb}}$ as a function of $t_g/t$ for various values of $U/t$.  Again, it is clear that there are two distinct regimes -- for $t_g$ comparable to $t$, there is a substantial value of $\Delta_{\text{pb}}$, of order $\Delta_s$, while for $t_g \lesssim 0.3t$, $\Delta_{\text{pb}}$ approaches 0.  (Given the limits on system sizes accessible to us, we are not confident that we can distinguish the difference between $\Delta_{\text{pb}} < 0.01t$ and $\Delta_{\text{pb}}=0$.)  These results are consistent with expectations based on the schematic phase diagram. Note the honeycom ladder spin gap regime at $U \sim 4t$ fits with recent studies on a suspected exotic magnetically disordered insulating state regime in the two-dimensional honeycomb Hubbard model~\cite{assaadnature}. Even though this claim turned out to be disproven by more accurate procedures~\cite{sorella,fakher2}, it is intuitive that we find such a spin gap in our quasi-1d honeycomb ladder scenario.

\section{DMRG results for the lightly doped honeycomb ladder}
\label{sec:nonhalf}
$H^{\text{eff}}$ in Eq.~\ref{Heff} encodes the properties of one or two electrons added to the undoped honeycomb ladder.  It is typical that the properties of the lightly doped system can be inferred from this kind of information, but there is always the chance that the behavior of a small  {\em concentration} $x$ of doped electrons is different from that of a small {\em number}.  
Consider, for instance, the spin-gap and the pair-binding energy as a function of  doped electron concentration:
$\Delta_{\text{pb}}(x)\equiv \lim_{L\to\infty} \Delta_{\text{pb}}(L,x)$ and $\Delta_{s}(x)\equiv \lim_{L\to\infty} \Delta_{s}(L,x)$ where (with the constraint that $2Lx$ is an even integer)
\bea
&\Delta_{\text{pb}}(L,x)&\equiv (1/2)\Big[2{\cal E}_L(2L+2Lx+1) \nonumber \\
&&-{\cal E}_L(2L+2Lx+2)-{\cal E}_L(2L+2Lx)\Big],
\\
&\Delta_{s}(L,x)&\equiv (1/2)\Big[2{\cal E}_L(2L+2Lx,S=1) \nonumber 
\\
&& -{\cal E}_L(2L+2Lx+2,S=0)\Big].
\eea
\begin{figure}[t]
\includegraphics[width=2.5in,angle=-90]{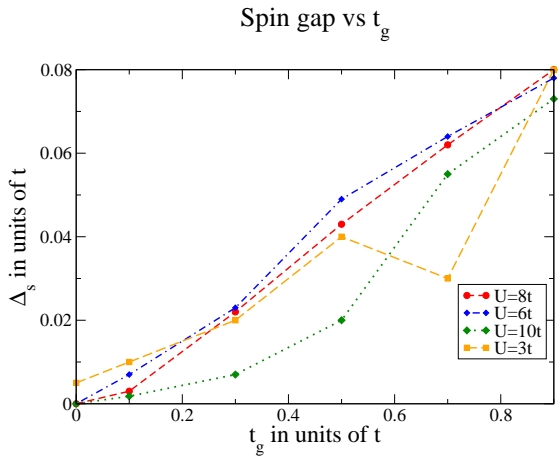}
\caption{
(Color online) Spin gap of the honeycomb ladder for various values of $U/t$  in the presence of two doped electrons ($x=1/L$), extrapolated to the thermodynamic limit $L\to\infty$.
For $t_g<0.3t$, 
the value of $\Delta_s$ is substantially reduced from its value at $x=0$ and zero within our numerical uncertainty.  By contrast, for $t_g\approx t$, 
the value of $\Delta_s$ is comparable to its value for $x=0$.}
\label{fig:2espingap}
\end{figure}
%
 While we did not carry these calculations out as extensively as for the undoped system, we have verified that for the 
 honeycomb ladder, $\Delta_{\text{pb}}$ and $\Delta_s$ both vanish (within our numerical uncertainty) for any small but non-zero $x$.  
 For example, 
 in the presence of two doped electrons ($x=1/L$), 
 the inferred values of $\Delta_s$ are only slightly smaller than for $x=0$ when $t_g=t$, but are reduced to values which are indistinguishable from 0 for $t_g < 0.3 t$, 
 as shown in Fig. \ref{fig:2espingap}.
 This is consistent with the notion that the lightly doped honeycomb ladder is describable in terms of dilute quasiparticles with the quantum numbers of an electron and with residual repulsive interactions.  This, in turn, is consistent with the effective Hamiltonian in Eq.~\ref {Heff}.

\section{Renormalization of the electron-phonon coupling}
\label{sec:phonon}

The electron-phonon vertex, in general, is renormalized in complex ways by electron-electron interactions resulting in a complicated function of the incoming electron and phonon momenta, which depends on the both the phonon and the electron band indices.  However, since the
 electron-phonon coupling is still typically rather local, we will study renormalization effects by explicitly computing the coupling to $k=0$ optical phonons for which $\delta t$ and $\delta\epsilon$ are periodic functions on the honeycomb lattice. 
 Operationally what this means is that we will compute changes in the electronic states of the system to linear order in these distortions as a function of  $U$.   Moreover, as we are interested in the case of lightly doped ladders, we are only concerned with how a phonon interacts with a doped electron or hole, {\it i.e.} how a phonon affects the quasiparticle energy, $\Delta_{\text{qp}}$.  We thus define the dimensionless electron phonon couplings:
\be
\beta_{ij} = N \frac {\partial \Delta_{\text{qp}}}{\partial\delta t_{ij}} \ \ \ {\rm and} \ \ \ \beta_j=N  \frac {\partial \Delta_{\text{qp}}}{\partial\delta \epsilon_j}.
\ee

There are five distinct nearest-neighbor bonds $(ij)$ per unit cell, but many of these are related by symmetry.  As a consequence, it is easy to see that  there are only two distinct values of $\beta_{ij}$:  $\beta_{\text{rung}}$ which is the response of the system to a change in the hopping matrix, $\delta t_{ij}$, on the nearest neighbor bond along a rung, and $\beta_{\text{leg}}$ for $\delta t_{ij}$ on one of the four bonds along the legs of the ladder.  Similarly, there are four distinct sites per unit cell, but the corresponding sites on the upper and lower leg of the ladder are related by symmetry, leaving two distinct coupling constants, $\beta_{\text{even}}$ and $\beta_{\text{odd}}$ corresponding to $\delta \epsilon_j$ with $j=(\alpha,n)$ and $n$ even ({\it i.e.} corresponding to a site at the apex of the elementary hexagon) or $n$ odd.  A uniform change in all site energies simply results in an additive contribution to $\Delta_{\text{qp}}$, from which it follows that
\be
\beta_{\text{even}}+\beta_{\text{odd}}=1.
\ee

\begin{figure}[t]
\includegraphics[width=2.5in,angle=-90]{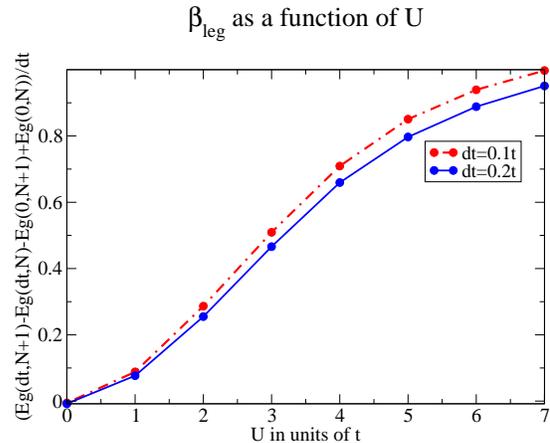}
\caption{
(Color online) Electron-phonon coupling on the legs:  Plotted is the discrete derivative $\beta_{\text{rung}}=\delta \Delta_{\text{qp}}/\delta t_{\text{leg}}$ for two values of $\delta t_{\text{leg}}$. 
Note that $\beta_{\text{leg}}\to 0$ in the non-interacting limit $U\to 0$.}
\label{fig:bleg}
\end{figure}

\begin{figure}[t]
\includegraphics[width=2.5in,angle=-90]{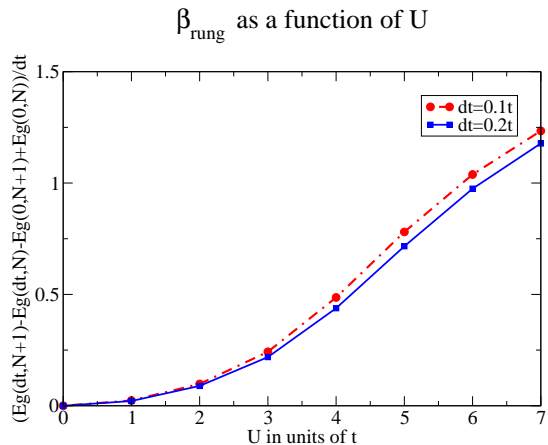}
\caption{
(Color online) Electron-phonon coupling on the rungs:  Plotted is the discrete derivative $\beta_{\text{rung}}=\delta \Delta_{\text{qp}}/\delta t_{\text{rung}}$ for two values of $\delta t_{\text{rung}}$. 
Note that $\beta_{\text{rung}}\to 0$ in the non-interacting limit $U\to 0$.}
\label{fig:brung}
\end{figure}

This leaves us with three dimensionless couplings to compute.  For $U=0$, it is easy to see that $\beta_{\text{rung}}=\beta_{\text{leg}}=\beta_{\text{odd}}=0$ and $\beta_{\text{even}}=1$.  The reason the first three couplings vanish for $U=0$ derives from the same special feature of the electronic structure responsible for the quadratic band touching in the honeycomb ladder.  Specifically, it is easily verified that at $k=\pi/2$, the Bloch states have a particularly simple structure:  Two bands are localized entirely on odd numbered sites and have energies $E(\pi/2) =-\lambda t$ and the other two are localized entirely on the even numbered sites and have energies $E(\pi/2) = -\lambda t_g$, where $\lambda=\pm 1$ labels the parity of the state under exchange of the legs.  For $t_g\to 0$, the latter two states constitute  the degenerate points of the quadratic band touching. Because these states are localized on the even numbered sites, they are (to first order) insensitive to any changes in hopping matrix elements (other than the third-neighbor coupling $t_g$, if we had included it), and moreover to changes in the site energies on the odd numbered sites.   

For non-zero $U/t$, we compute the various couplings using DMRG by introducing an appropriate set  small changes $\delta t_{ij}$ or $\delta \epsilon_i$ in each unit cell, computing the change in $\Delta_{\text{qp}}$ and dividing by  
$\delta t$ or $\delta \epsilon$.  By doing this for a couple of different magnitudes of the change, we confirm that we are obtaining the derivative.  The results for $\beta_{\text{leg}}$ and $\beta_{\text{rung}}$ as a function of $U/t$ are shown in Fig.~\ref{fig:bleg} and Fig.~\ref{fig:brung}, respectively. 
Notice that in the presence of  reasonable values of $U/t$,
 $\beta_{\text{leg}}$ and $\beta_{\text{rung}}$ are substantial;  there is, in this sense, an infinite renormalization of the electron-phonon coupling produced by the interactions.

 
 Experience with other conducting polymers, such as polyacetylene, suggests that the dominant electron-phonon coupling is associated with modulation of the hopping matrix elements in response to changes in the bond lengths.  This dominant form of coupling vanishes identically in the honeycomb ladder with $U=0$.  This is why the renormalization due to $U$ is so important -- it produces non-zero coupling to the important phonons.  In terms of the quasiparticle creation operators that appear in Eq.~\ref{Heff}, the effective electron-phonon coupling can be expressed as
 \bea
&H_{\text{el-ph}}^{\text{eff}}& = \sum_R\Big[\beta_{\text{rung}}\delta t_{R}^{(\text{rung})} +\beta_{\text{leg}}\delta t_{R}^{(\text{leg})} \nonumber\\
&&+\beta_{\text{even}}\delta\epsilon_{R}^{(\text{even})}+\beta_{\text{odd}}\delta\epsilon_{R}^{(\text{odd})}\Big ] \rho(R),
 \eea
 where $\rho(R) =\sum_\sigma \Psi^\dagger_{R\sigma}\Psi_{R\sigma}$, $\delta t_{R}^{(leg)}$ is defined to be the sum over all four legwise bonds in unit cell $R$ of $\delta t_{ij}$ and similarly for the remaining terms, where implicitly $\delta\epsilon_R^{(a)}$ and $\delta t_{\text{F}}^{(a)}$ are functions of the phonon coordinates.

We have not attempted to treat the full electron-phonon problem, as there are many modes and, moreover, the phonon dispersion would certainly depend on details of the solid state environment in any physical world realization of the present model.  We thus simply estimate the strength of the resulting effective attraction as
\be
V^{\text{eff}} \approx 5 \beta^2 \alpha^2/K \approx \beta^2 4 eV,
\ee
where $\beta$ is an appropriate average of $\beta_{\text{rung}}$ and $\beta_{\text{leg}}$, $\alpha$ is an appropriate average of $\alpha_{ij}\equiv\partial t_{ij}/\partial u_{ij}$ where $u_{ij}$ is the length of the bond $(ij)$, $K$ is an average spring-constant and the factor of 5 comes from the number of bonds per unit cell;  the second equality uses empirical estimates of the corresponding quantities in polyacetylene, $\alpha=3.6 $eV$\AA^{-1}$ and $K=16$eV$\AA^{-2}$.\footnote{These couplings are generally reported in terms of coupling to displacements projected along the polymer chain -- in obtaining the above estimates, a geometric factor has been included to express the same quantities in terms of changes in the bond-lengths.}

\section{Discussion: Superconductivity in lightly doped polyacenes}
\label{sec:discussion}
The interactions in any polyacene-based material are certainly more complex than in the honeycomb Hubbard ladder.  Nonetheless, we feel that the basic results from our studies are sufficiently robust in that they should apply to long enough polymers in the solid state.  Specifically:

\noindent{\bf 1)} The undoped polymer is expected to be a robust Mott insulator, with a spin-gap small compared to its charge gap, even for relatively weak interactions.

\noindent{\bf 2)} Despite the presence of low energy spin fluctuations, the lightly doped polymer will not
 exhibit unconventional superconductivity mediated by spin fluctuations, due to the absence of pair binding
 energy, even in the lightly doped polymers. (Pair-binding has been found 
robustly in a large variety of ladder systems, so its absence in the honeycomb ladder is somewhat notable.)

\noindent{\bf 3)} Lightly doped polymeric solids, however, are prime candidates for high temperature conventional superconductivity.  The dominant phonons in this are likely to be those that modulate the hopping matrix elements, $t_{ij}$, which couple to the quasiparticles through an interaction-induced coupling.   Our estimates suggest that the phonon-induced effective attraction, $V^{\text{eff}}\sim 1$eV for a characteristic value\cite{RMP} of $U=4t$, is greater than the residual effective repulsion, $U^{\text{eff}}\sim$ 0.1eV, and thus that a mean-field superconducting transition is indicated.

\noindent{\bf 4)} We have discovered  significant correlation enhancements of the electron-phonon couplings:  For $U=0$, the majority of phonons do not couple linearly to the band edge states.     
However, as shown in Figs.~\ref{fig:brung} and~\ref{fig:bleg}, these couplings grow in proportion to $N(E_{\text{F}})U^2$.  This novel correlation effect, combined with the characteristic 1D divergence of the density of states at low doping, $N(E_{\text{F}}) \sim |x|^{-1/2}$, implies that $\lambda$ is substantial in relatively lightly doped polyacenes.  

\noindent{\bf 5)} The lower the doping concentration, the higher the density of states, which is conducive to a higher pairing scale.  However, at very low doping, phase fluctuations -- always a significant issue in quasi 1D systems -- are likely to suppress $T_c$.  Thus, optimal doping concentration for superconductivity is likely to occur when $E_{\text{F}} \geq \hbar \omega_0$, {\it i.e.} when $ x=x_{max} \geq \sqrt{\hbar\omega_0/4\pi^2t}$.

\begin{acknowledgements}
We acknowledge helpful discussions with E. Berg, L. Forro, and R. C. Haddon. This work was supported in part by DOE grant No AC02-76SF00515 at Stanford (G.K. and S.A.K.) and by DFG-SPP 1458 and ERC-StG-2013-336012 (R.T.). 
\end{acknowledgements}

\appendix

\section{Graphene ribbons and modified graphite}
\label{app:graphene}

\begin{figure}[t!]
\includegraphics[width=3.2in]{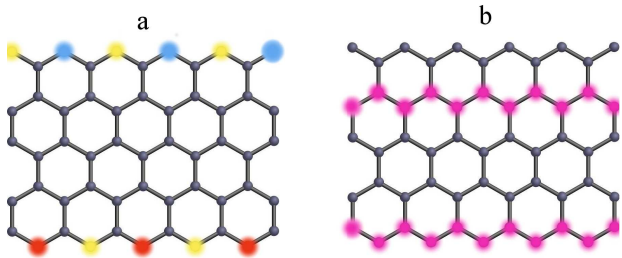}
\caption{(Color online) a) Graphene strips support zero energy edge states, represented by the positions 
in which the sign of the wavefunction alternates (blue dots for $+$ and yellow for $-$ for the wavefunction associated with the upper end, red dots for $+$ and yellow for $-$ in the case of the lower end.)
b) In this configuration we suggest every third zig-zag chain (horizontal orientation) to exhibit different
on site energies (purple dots), and as a consequence, to exhibit quasi-1d ladder behavior in between. }
\label{fig:graphene}
\end{figure}

The unique aspects of the electronic structure of the honeycomb ladder depend on its geometry rather than details of orbital chemistry.  Thus, aspects of the physics we have explored here may be relevant to a broader range of structures.  Here we discuss a few representative generalized structures.

Firstly, we consider the case of a broader graphene strip, as shown schematically in Fig.~\ref{fig:graphene}a, and for simplicity we discuss the case in which there are non-zero hopping matrix elements only between nearest-neighbor sites.  Notably, this system also has a quadratic band touching as can be seen by the following analysis:  1)  At zero energy, there are two degenerate states with $k=\pi/2$.  One is a state which lives on the upper row of sites and alternates in sign along the stripe, as represented by the white and blue spheres in the Fig.~\ref{fig:graphene}a.  Of course, there is also a corresponding zero mode on the lower edge, as represented by the alternating white and red spheres in Fig.~\ref{fig:graphene}.  In a graphene strip of large width, these states would be part of non-dispersing edge states which, for a range of $\pi/3 < k < 2\pi/3$, are confined within a distance $\lambda_k = -1/\ln|2\cos(k)|$ of the edge,  and which mix with the bulk nodal states when $k=\pi/3$ or $-2\pi/3$.  For a finite width strip, the mixing of these states produces a higher order band touching  $k=\pi/2$. 
 
If the strip is much wider than the ladders we have considered, the coupling between the edge states, and with it their dispersion, is negligible.  Such flat bands will more likely lead to some form of ferromagnetic order rather than to superconductivity.  
However, a more promising possibility is to consider a decorated version of graphene (or graphite) with a structure such as that shown schematically in Fig.~\ref{fig:graphene}b.  There, we have imagined that in every third zig-zag row of the honeycomb lattice, there is a modification to the site which to a large part removes it from the electronic structure.  For instance, one might imagine (if such a material could be made) that on these sites, the C atoms are replaced with N atoms, with one extra electron and a larger ionic charge.  Such a material would have an electronic structure near the Fermi energy which can be thought of as arising from weak coupling between an oriented array of honeycomb ladders.  Undoped,  on the basis of our ladder resultsj, such a material would be expected to form a narrow gap quasi-1D semiconductor.  When doped, if luck is with us, it could be a high temperature electron-phonon driven superconductor.

\bibliographystyle{apsrev}
\bibliography{bibliography}

\begin{thebibliography}{17}
\expandafter\ifx\csname natexlab\endcsname\relax\def\natexlab#1{#1}\fi
\expandafter\ifx\csname bibnamefont\endcsname\relax
  \def\bibnamefont#1{#1}\fi
\expandafter\ifx\csname bibfnamefont\endcsname\relax
  \def\bibfnamefont#1{#1}\fi
\expandafter\ifx\csname citenamefont\endcsname\relax
  \def\citenamefont#1{#1}\fi
\expandafter\ifx\csname url\endcsname\relax
  \def\url#1{\texttt{#1}}\fi
\expandafter\ifx\csname urlprefix\endcsname\relax\def\urlprefix{URL }\fi
\providecommand{\bibinfo}[2]{#2}
\providecommand{\eprint}[2][]{\url{#2}}

\bibitem[{\citenamefont{Bardeen et~al.}(1957)\citenamefont{Bardeen, Cooper, and
  Schrieffer}}]{bcs}
\bibinfo{author}{\bibfnamefont{J.}~\bibnamefont{Bardeen}},
  \bibinfo{author}{\bibfnamefont{L.~N.} \bibnamefont{Cooper}},
  \bibnamefont{and} \bibinfo{author}{\bibfnamefont{J.~R.}
  \bibnamefont{Schrieffer}}, \bibinfo{journal}{Phys. Rev.}
  \textbf{\bibinfo{volume}{108}}, \bibinfo{pages}{1175} (\bibinfo{year}{1957}),
  \urlprefix\url{http://link.aps.org/doi/10.1103/PhysRev.108.1175}.

\bibitem[{\citenamefont{Kivelson and Chapman}(1983)}]{Stevechap}
\bibinfo{author}{\bibfnamefont{S.}~\bibnamefont{Kivelson}} \bibnamefont{and}
  \bibinfo{author}{\bibfnamefont{O.~L.} \bibnamefont{Chapman}},
  \bibinfo{journal}{Phys. Rev. B} \textbf{\bibinfo{volume}{28}},
  \bibinfo{pages}{7236} (\bibinfo{year}{1983}),
  \urlprefix\url{http://link.aps.org/doi/10.1103/PhysRevB.28.7236}.

\bibitem[{\citenamefont{Su et~al.}(1980)\citenamefont{Su, Schrieffer, and
  Heeger}}]{ssh}
\bibinfo{author}{\bibfnamefont{W.~P.} \bibnamefont{Su}},
  \bibinfo{author}{\bibfnamefont{J.~R.} \bibnamefont{Schrieffer}},
  \bibnamefont{and} \bibinfo{author}{\bibfnamefont{A.~J.}
  \bibnamefont{Heeger}}, \bibinfo{journal}{Phys. Rev. B}
  \textbf{\bibinfo{volume}{22}}, \bibinfo{pages}{2099} (\bibinfo{year}{1980}),
  \urlprefix\url{http://link.aps.org/doi/10.1103/PhysRevB.22.2099}.

\bibitem[{\citenamefont{Xue et~al.}(2011)\citenamefont{Xue, T., Wang, Wu, Yang,
  He, Li, and Chen}}]{xue}
\bibinfo{author}{\bibfnamefont{M.}~\bibnamefont{Xue}},
  \bibinfo{author}{\bibfnamefont{C.}~\bibnamefont{T.}},
  \bibinfo{author}{\bibfnamefont{D.}~\bibnamefont{Wang}},
  \bibinfo{author}{\bibfnamefont{Y.}~\bibnamefont{Wu}},
  \bibinfo{author}{\bibfnamefont{X.}~\bibnamefont{Yang},
  \bibfnamefont{H.~Dong}},
  \bibinfo{author}{\bibfnamefont{J.}~\bibnamefont{He}},
  \bibinfo{author}{\bibfnamefont{F.}~\bibnamefont{Li}}, \bibnamefont{and}
  \bibinfo{author}{\bibfnamefont{G.~F.} \bibnamefont{Chen}},
  \bibinfo{journal}{Scientific Reports} \textbf{\bibinfo{volume}{2}},
  \bibinfo{pages}{389} (\bibinfo{year}{2011}).

\bibitem[{\citenamefont{Chun et~al.}(2008)\citenamefont{Chun, Cheng, and
  Wudl}}]{heptacene}
\bibinfo{author}{\bibfnamefont{D.}~\bibnamefont{Chun}},
  \bibinfo{author}{\bibfnamefont{Y.}~\bibnamefont{Cheng}}, \bibnamefont{and}
  \bibinfo{author}{\bibfnamefont{F.}~\bibnamefont{Wudl}},
  \bibinfo{journal}{Angew. Chem. Int. Ed.} \textbf{\bibinfo{volume}{47}},
  \bibinfo{pages}{8380} (\bibinfo{year}{2008}).

\bibitem[{\citenamefont{White}(1992)}]{white1}
\bibinfo{author}{\bibfnamefont{S.~R.} \bibnamefont{White}},
  \bibinfo{journal}{Phys. Rev. Lett.} \textbf{\bibinfo{volume}{69}},
  \bibinfo{pages}{2863} (\bibinfo{year}{1992}),
  \urlprefix\url{http://link.aps.org/doi/10.1103/PhysRevLett.69.2863}.

\bibitem[{\citenamefont{White}(1993)}]{white2}
\bibinfo{author}{\bibfnamefont{S.~R.} \bibnamefont{White}},
  \bibinfo{journal}{Phys. Rev. B} \textbf{\bibinfo{volume}{48}},
  \bibinfo{pages}{10345} (\bibinfo{year}{1993}),
  \urlprefix\url{http://link.aps.org/doi/10.1103/PhysRevB.48.10345}.

\bibitem[{\citenamefont{Schollw\"ock}(2005)}]{scholl}
\bibinfo{author}{\bibfnamefont{U.}~\bibnamefont{Schollw\"ock}},
  \bibinfo{journal}{Rev. Mod. Phys.} \textbf{\bibinfo{volume}{77}},
  \bibinfo{pages}{259} (\bibinfo{year}{2005}),
  \urlprefix\url{http://link.aps.org/doi/10.1103/RevModPhys.77.259}.

\bibitem[{\citenamefont{Noack et~al.}(1992)\citenamefont{Noack, White, and
  Scalapino}}]{2LegHubbard}
\bibinfo{author}{\bibfnamefont{R.}~\bibnamefont{Noack}},
  \bibinfo{author}{\bibfnamefont{S.}~\bibnamefont{White}}, \bibnamefont{and}
  \bibinfo{author}{\bibfnamefont{D.}~\bibnamefont{Scalapino}},
  \bibinfo{journal}{Physica C} \textbf{\bibinfo{volume}{270}},
  \bibinfo{pages}{281} (\bibinfo{year}{1992}).

\bibitem[{\citenamefont{Balents and Fisher}(1996)}]{balentsandfisher}
\bibinfo{author}{\bibfnamefont{L.}~\bibnamefont{Balents}} \bibnamefont{and}
  \bibinfo{author}{\bibfnamefont{M.~P.~A.} \bibnamefont{Fisher}},
  \bibinfo{journal}{Phys. Rev. B} \textbf{\bibinfo{volume}{53}},
  \bibinfo{pages}{12133} (\bibinfo{year}{1996}).

\bibitem[{\citenamefont{Karakonstantakis
  et~al.}(2011)\citenamefont{Karakonstantakis, Berg, White, and
  Kivelson}}]{george1}
\bibinfo{author}{\bibfnamefont{G.}~\bibnamefont{Karakonstantakis}},
  \bibinfo{author}{\bibfnamefont{E.}~\bibnamefont{Berg}},
  \bibinfo{author}{\bibfnamefont{S.~R.} \bibnamefont{White}}, \bibnamefont{and}
  \bibinfo{author}{\bibfnamefont{S.~A.} \bibnamefont{Kivelson}},
  \bibinfo{journal}{Phys. Rev. B} \textbf{\bibinfo{volume}{83}},
  \bibinfo{pages}{054508} (\bibinfo{year}{2011}),
  \urlprefix\url{http://link.aps.org/doi/10.1103/PhysRevB.83.054508}.

\bibitem[{\citenamefont{Zou et~al.}(2011)\citenamefont{Zou, Hong, and
  Zhu}}]{graphene}
\bibinfo{author}{\bibfnamefont{K.}~\bibnamefont{Zou}},
  \bibinfo{author}{\bibfnamefont{X.}~\bibnamefont{Hong}}, \bibnamefont{and}
  \bibinfo{author}{\bibfnamefont{J.}~\bibnamefont{Zhu}},
  \bibinfo{journal}{Phys. Rev. B} \textbf{\bibinfo{volume}{84}},
  \bibinfo{pages}{085408} (\bibinfo{year}{2011}),
  \urlprefix\url{http://link.aps.org/doi/10.1103/PhysRevB.84.085408}.

\bibitem[{\citenamefont{Meng et~al.}(2010)\citenamefont{Meng, Lang, Wessel,
  Assaad, and Muramatsu}}]{assaadnature}
\bibinfo{author}{\bibfnamefont{Z.~Y.} \bibnamefont{Meng}},
  \bibinfo{author}{\bibfnamefont{T.~C.} \bibnamefont{Lang}},
  \bibinfo{author}{\bibfnamefont{S.}~\bibnamefont{Wessel}},
  \bibinfo{author}{\bibfnamefont{F.~F.} \bibnamefont{Assaad}},
  \bibnamefont{and}
  \bibinfo{author}{\bibfnamefont{A.}~\bibnamefont{Muramatsu}},
  \bibinfo{journal}{Nature} \textbf{\bibinfo{volume}{464}},
  \bibinfo{pages}{847} (\bibinfo{year}{2010}).

\bibitem[{\citenamefont{Sorella et~al.}(2012)\citenamefont{Sorella, Otsuka, and
  Yunoki}}]{sorella}
\bibinfo{author}{\bibfnamefont{S.}~\bibnamefont{Sorella}},
  \bibinfo{author}{\bibfnamefont{Y.}~\bibnamefont{Otsuka}}, \bibnamefont{and}
  \bibinfo{author}{\bibfnamefont{S.}~\bibnamefont{Yunoki}},
  \bibinfo{journal}{Scientific Reports} \textbf{\bibinfo{volume}{2}},
  \bibinfo{pages}{992} (\bibinfo{year}{2012}).

\bibitem[{\citenamefont{Assaad and Herbut}()}]{fakher2}
\bibinfo{author}{\bibfnamefont{F.~F.} \bibnamefont{Assaad}} \bibnamefont{and}
  \bibinfo{author}{\bibfnamefont{I.~F.} \bibnamefont{Herbut}},
  \bibinfo{note}{arXiv:1304.6340}.

\bibitem[{Note1()}]{Note1}
Note1, \bibinfo{note}{these couplings are generally reported in terms of
  coupling to displacements projected along the polymer chain -- in obtaining
  the above estimates, a geometric factor has been included to express the same
  quantities in terms of changes in the bond-lengths.}

\bibitem[{\citenamefont{Heeger et~al.}(1988)\citenamefont{Heeger, Kivelson,
  Schrieffer, and Su}}]{RMP}
\bibinfo{author}{\bibfnamefont{A.~J.} \bibnamefont{Heeger}},
  \bibinfo{author}{\bibfnamefont{S.}~\bibnamefont{Kivelson}},
  \bibinfo{author}{\bibfnamefont{J.~R.} \bibnamefont{Schrieffer}},
  \bibnamefont{and} \bibinfo{author}{\bibfnamefont{W.~P.} \bibnamefont{Su}},
  \bibinfo{journal}{Rev. Mod. Phys.} \textbf{\bibinfo{volume}{60}},
  \bibinfo{pages}{781} (\bibinfo{year}{1988}),
  \urlprefix\url{http://link.aps.org/doi/10.1103/RevModPhys.60.781}.

\end{thebibliography}

\end{document}